\documentclass[12pt, a4paper]{article}
\pdfoutput=1

\usepackage{amsmath}
\usepackage{amsfonts}
\usepackage{amssymb}
\usepackage{graphicx, rotating}
\usepackage{epsfig}
\usepackage{latexsym}
\usepackage{graphicx}
\usepackage{color}
\usepackage{amsmath,bm,amssymb}
\usepackage{cite}
\usepackage{slashed}
\usepackage{hyperref}
\hypersetup{colorlinks, citecolor=bluscuro, linkcolor=black, urlcolor=bluscuro}
\definecolor{rossos}{cmyk}{0,1,1,0.55}
\definecolor{bluscuro}{rgb}{0.15, 0.2, .85}
\definecolor{bluchiaro}{cmyk}{1,.3,0.,0.1}

\newcommand{\be}{\begin{equation}}
\newcommand{\ee}{\end{equation}}
\newcommand{\bea}{\begin{eqnarray}}
\newcommand{\eea}{\end{eqnarray}}

\newcommand{\arXiv}[2]{\href{http://arxiv.org/pdf/#1}{{\tt [#2/#1]}}}
\newcommand{\arXivold}[1]{\href{http://arxiv.org/pdf/#1}{{\tt [#1]}}}

\def\bma#1{\mbox{\boldmath{$#1$}}}

\begin{document}
\allowdisplaybreaks
\begin{titlepage}
\begin{flushright}
DESY-23-223 
\end{flushright}
\vspace{.3in}

\vspace{1cm}
\begin{center}
{\Large\bf\color{black} 
Exact Tunneling Solutions in Multi-Field Potentials} \\
\vspace{1cm}{
{\large J.R.~Espinosa$^a$, T.~Konstandin$^b$}
\vspace{0.3cm}
} \\[7mm]
{\it {$^a$\,  Instituto de F\'{\i}sica Te\'orica, IFT-UAM/CSIC, \\ 
C/ Nicol\'as Cabrera 13-15, Campus de Cantoblanco, 28049, Madrid, Spain}}\\
{\it $^b$ {Deutsches Elektronen-Synchrotron DESY, Notkestr. 85, 22607 Hamburg, Germany}}
\end{center}
\bigskip

\vspace{.4cm}

\begin{abstract}

The tunneling potential formalism makes it easy to construct exact solutions to the vacuum decay problem in potentials with multiple fields. 
While some exact solutions for single-field decays were known, we present the first nontrivial analytic examples with two and three scalar fields, and show how the method can be generalized to include gravitational corrections. 
Our results illuminate some analytic properties of the tunneling potential functions and can have a number of uses, among others: to serve as simple approximations to realistic potentials; to learn about parametric dependencies of decay rates; to check conjectures on vacuum decay; 
as benchmarks for multi-field  numerical codes; or to study holographic interpretations of vacuum decay.

\end{abstract}
\bigskip

\end{titlepage}

\section{Introduction \label{sec:intro}} 

Vacuum decay via quantum tunneling is a ubiquitous phenomenon which is relevant in many areas of particle physics and cosmology (as well as condensed matter systems). A cursory look at the references to the pioneering work of Coleman \cite{Coleman} (without gravity) and Coleman and De Luccia \cite{CdL} (with gravity) corroborates this.  There is a vast range of physical systems where decay of metastable states is of central importance, from cosmological phase transitions to particle physics scenarios beyond the standard model featuring multiple vacua; from the string landscape to the AdS/CFT correspondence, etc. 
Exact solutions to the tunneling problem (beyond the thin-wall limit) are useful for several reasons: 

{\it 1)} They can be good approximations to more complicated situations and allow a parametric understanding of the real problem. One relevant example is the (negative) quartic potential $V=-(\lambda/4)\varphi^4$ which is a simple approximation to the potential of the Higgs  field at field values much larger than the electroweak scale (see {\it e.g.} \cite{SMinst}). This potential admits an exact tunneling solution (the Fubini instanton, in the Euclidean formulation of the decay problem) which  allows to calculate the tunneling action for decay, $S=8\pi^2/(3\lambda)$.

{\it 2)} They can be used to learn about the analytic structure and parametric dependencies of decay rates. As one particular example, such analytically solvable potentials were used in \cite{Paban} to study how the tunneling action scales with the size and widths of barriers both for single-field potentials and additive multi-field potentials.  Another example can be found in \cite{ESM}, which explores analytically the parametric dependence of the tunneling action for the decay of the Standard Model vacuum including running of $\lambda$, gravitational corrections and a non-minimal coupling of the Higgs field to gravity.

{\it 3)} They can be used to check (and discard) conjectures about some type of vacuum decay. For example, the negative quartic potential discussed in {\it 1)} above admits a family of tunneling instantons with degenerate action as a result of the scale invariance of the potential. Thus, one might conjecture that scale invariance is necessary to have such action degeneracy. However, \cite{BoN} found a simple analytic counterexample to this (see subsection~7.5 of that paper).

{\it 4)} They can serve as testbeds for numerical codes \cite{VEVACIOUS,CT,AB,BP,SB,FB} that are often employed in practice, particularly for the study of multi-field potentials, for which one cannot rely on the overshoot/undershoot method to find solutions.

{\it 5)} For the case with gravity, vacuum decay is associated to Coleman-De Luccia (CdL) geometries \cite{CdL}. Analytic CdL geometries beyond thin-wall can be useful \cite{DH} to study the connection between bubble nucleation in eternal inflation and a Euclidean conformal field theory via a holographic duality \cite{holo}. Similarly, such examples can also be useful to study in detail Maldacena's interpretation \cite{AdS} of decays of a false vacuum into an AdS region in terms of a dual field theory living on an end of the world brane expanding into the false vacuum.

The aim of this paper is to provide a constructive method to generate exact solutions starting from the single-field case. 
The main tool is  the tunneling potential method that provides an alternative description to tunneling~\cite{E}
and takes the equations to a particularly simple form in the case with gravity \cite{Eg}.
 
\section{Exact Tunneling Solutions for a Single Field \label{sec:field}} 

The tunneling potential method poses the determination of the tunneling action for the decay of a false vacuum of $V(\varphi)$ at $\varphi_+$  in the following form: find the (tunneling potential) function $V_t(\varphi)$, which goes from $\varphi_+$ to some $\varphi_0$ on the basin of the true vacuum at $\varphi_-$, and minimizes the action functional \cite{E}
\be
 S[V_t]=54\pi^2\int_{\varphi_+}^{\varphi_0} \frac{(V-V_t)^2}{(-V'_t)^3}\ d\varphi\ ,
\label{SVt}
\ee
where we assume $\varphi_+<\varphi_0< \varphi_-$ and define $x'\equiv dx/d\varphi$. The method reproduces the Euclidean bounce result \cite{Coleman} and has a number of good properties discussed elsewhere. 
The  Euler-Lagrange equation, $\delta S/\delta V_t=0$, gives the``equation of motion'' (EoM) for $V_t$ as
\be
(4V_t'-3V')V_t' + 6(V-V_t)V_t''=0
\ .
\label{EoM}
\ee 
The boundary conditions for the tunneling solution are
\be
V_t(\varphi_+)=V(\varphi_+)\ ,\quad V_t'(\varphi_+)=V'(\varphi_+)=0\ , \quad
V_t(\varphi_0)=V(\varphi_0)\ ,\quad V_t'(\varphi_0)=\frac34 V'(\varphi_0)\ .
\ee
Without loss of generality we can set $\varphi_+=0$ and $V(\varphi_+)=0$.

To get exact tunneling solutions for a single scalar field $\varphi$ we solve the equation
\be
V(\varphi) = V_t(\varphi)+\frac{V_t'(\varphi)^2}{3}\int_{\varphi_0}^\varphi\frac{d\bar\varphi}{V_t'(\bar\varphi)}\ ,
\label{eq:sol_by_int}
\ee
where $V_t(\varphi)$ is a tunneling function simple enough for the integral to be solvable. The function $V_t$ is a monotonically decreasing function that connects the false vacuum at $\varphi=0$ to
a point $\varphi_0$ in the basin of the true vacuum of $V$. It can be checked that (\ref{eq:sol_by_int}) is indeed a solution to the tunneling potential EoM (\ref{EoM}).
The integration constant is fixed  by the condition $V(0)=V_t(0)$.

If the integral in (\ref{eq:sol_by_int}) can be performed, one can also obtain analytically the field profile, in the inverse form $r(\varphi)$. This follows from the relation $r^2=18(V-V_t)/V_t'{}^2$ \cite{E} which, used in (\ref{eq:sol_by_int}), gives
\be
r^2(\varphi)=6\int_{\varphi_0}^\varphi\frac{d\bar\varphi}{V_t'(\bar\varphi)}\ .
\label{r2phi}
\ee
In some cases this relation can be inverted to get $\varphi(r)$ explicitly, as we will show below, but having (\ref{r2phi}) analytically is equally useful. In terms of $r^2(\varphi)$ we can rewrite the potential (\ref{eq:sol_by_int}) as
\be
V(\varphi) = V_t(\varphi)+\frac{1}{18} V_t'{}^2(\varphi)r^2(\varphi)\ .
\label{Vr2}
\ee
We show below some analytic examples for $V(\varphi)$ and the reader can produce new ones easily. Previous exact solutions with a single field, obtained via different methods, can be found in \cite{exact}.

\subsection{Example A: Polynomial $\bma{V_t}$\label{sec:ExPolynomial}}

The simple monotonic tunneling potential function \cite{E}
\be
V_t = \varphi^2(2\varphi-3)\ ,
\label{eq:ex1}
\ee
leads to 
\be
V = \varphi^2\left[2\varphi-3+(1-\varphi)^2\log\frac{(1-\varphi)^2\varphi_0^2}{(1-\varphi_0)^2\varphi^2}\right]\ ,
\ee
with $0<\varphi_0<1$.
The tunneling action is
\be
S = -\frac{\pi^2}{3}\left[\varphi_0+{\rm Li}_2\left(\frac{\varphi_0}{\varphi_0-1}\right)\right]\ ,
\ee
where ${\rm Li}_2(x)$ is the dilogarithm function,
and the bounce profile is obtained as
\be
\varphi_B(r)=\frac{\varphi_0}{\varphi_0+(1-\varphi_0)e^{r^2}}=\frac{1-2 e^{-R^2}}{1-2 e^{-R^2}+e^{r^2-R^2}}\ ,
\ee
where $R$ is the radius of the bounce, defined by $\varphi_B(R)=\varphi_0/2$. This example has a thin-wall limit for large radius $R$ with $\varphi_0$ getting exponentially close to 1 (where $V_t$ has a minimum). The setup will reach the thin-wall limit by producing 
larger and larger barriers once the scalar potential $V$ is reconstructed using (\ref{eq:sol_by_int}). 
\begin{figure}[t!]
\begin{center}
\includegraphics[width=0.45\textwidth]{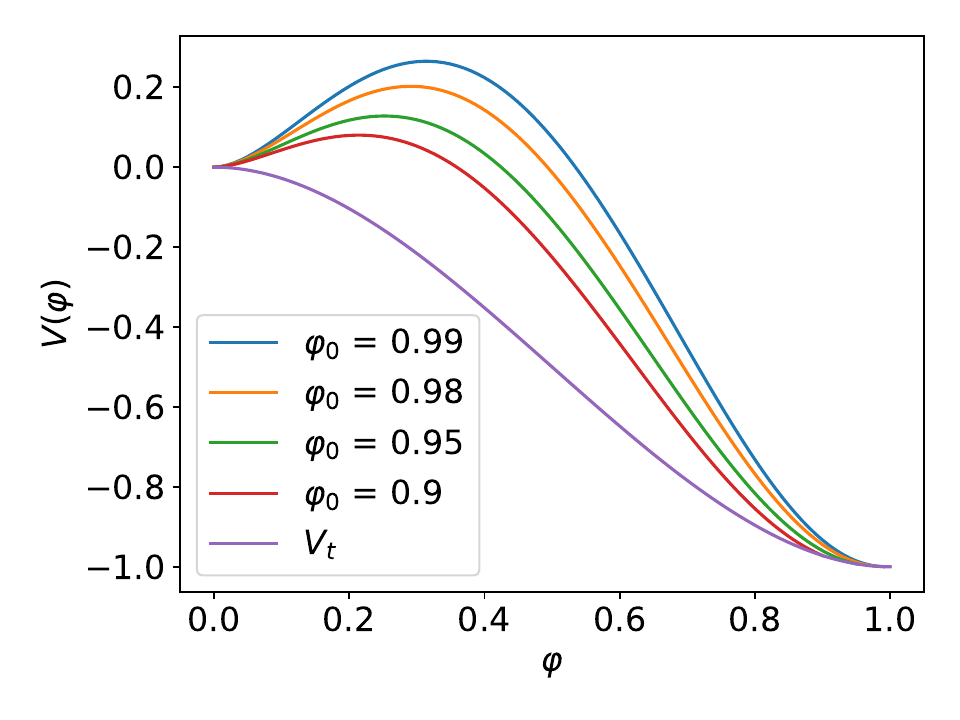}
\includegraphics[width=0.45\textwidth]{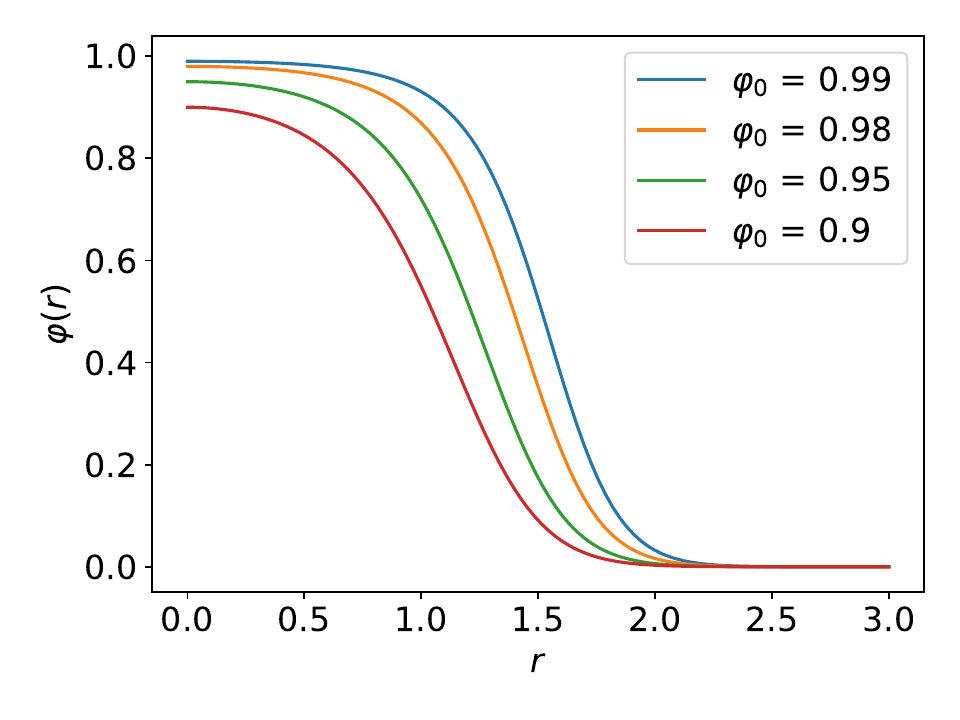}
\end{center}
\caption{
\label{fig:1field}
For the example of subsection~\ref{sec:ExPolynomial} and for several values of $\varphi_0$, potentials $V(\varphi)$ (left plot) and corresponding Euclidean bounce profiles (right plot).
}
\end{figure}
This is shown in Figure~\ref{fig:1field}.

A small drawback of this example is that the  small-field expansion of $V(\varphi)$ contains a $\varphi^2\log\varphi$ term, so that  the mass at the false vacuum is not properly defined. Although this is not a problem to test numerical codes, some examples below remedy this.

\subsection{Example B: Trigonometric tunneling potential}
The tunneling potential
\be
V_t = -\sin^2\varphi\ ,
\ee
gives 
\be
V = \left[-1+\frac23\cos^2\varphi\log\left(\frac{\tan\varphi_0}{\tan\varphi}\right)\right]\sin^2\varphi
\ ,
\ee
with $0<\varphi_0<\pi/2$.
The tunneling action is
\be
S =\frac{\pi^2}{4}\left[\frac{\pi^2}{2}+6\log^2(\cot\varphi_0)+3{\rm Li}_2(-\cot^2\varphi_0)\right] \ ,
\ee
and the bounce profile reads
\be
\varphi_B(r)={\rm arcot}\left(e^{r^2/3}\cot\varphi_0\right)={\rm arcot}\left(\frac{e^{r^2/3-R^2/6}}{\sqrt{e^{R^2/3}-2}}\right)\ ,
\ee
where $R$ is the radius of the bounce, defined by $\varphi_B(R)=\varphi_0/2$ and given in terms of $\varphi_0$ by $R^2=3\log(1+{\rm sec}\varphi_0)$. This example also has a thin-wall limit for large radius $R$ (with $\varphi_0\to\pi/2$). As in the previous example, the  small field expansion of $V(\varphi)$ contains a $\varphi^2\log\varphi$ term.

\subsection{Example C: Finite mass}
For
\be
V_t(\varphi)=\frac{\varphi^2}{-1/2+\log\varphi}\ ,
\ee
we obtain
\be
V(\varphi)=\frac{\varphi^2}{-1/2+\log\varphi}+\frac{8\varphi^2(1-\log\varphi)^2}{3(1-2\log\varphi)^2}\left(2\log^2\varphi-2\log^2\varphi_0+\log\frac{1-\log\varphi}{1-\log\varphi_0}\right)\ ,
\label{Vex2}
\ee
with $0<\varphi_0<\sqrt{e}$.
It can be checked that $V(\varphi)=\varphi^2/3+{\cal O}(\varphi^2/\log\varphi)$ so that the mass at the false vacuum is well defined.

In this case the tunneling action cannot be calculated analytically but it can be simply calculated numerically. The field profile $\varphi(r)$ cannot be obtained analytically either, but its inverse function is
\be
r^2(\varphi) =\frac34\left(2\log^2\varphi-2\log^2\varphi_0+\log\frac{1-\log\varphi}{1-\log\varphi_0}\right)\ ,
\ee
as can be read off from (\ref{Vex2}) using  (\ref{Vr2}).
This example does not admit a thin-wall limit though, since the tunneling potential does not feature a 
broken minimum.

\subsection{Example D: Derivative of the tunneling potential}

Another class of exact solutions can be obtained by starting from the derivative of the tunneling potential.
For
\be
V_t'(\varphi)=\frac{\varphi}{\log\varphi}\ ,
\ee
we get
\be
V_t(\varphi)={\rm Ei}(2\log\varphi)\ ,
\ee
where ${\rm Ei}(x)$ is the exponential integral function, which leads to
\be
V(\varphi)={\rm Ei}(2\log\varphi)+\frac16\varphi^2\left(1-\frac{\log^2\varphi_0}{\log^2\varphi}\right)\ ,
\ee
with $0<\varphi_0<1$.
This potential has the small-field expansion $V(\varphi)=\varphi^2/6+{\cal O}(\varphi^2/\log\varphi)$, so that the mass at the false vacuum is well defined.
In this case the tunneling action can be obtained as
\be
S=\frac{3\pi^2}{16}\left[-8\ {\rm Ei}(2\log\varphi_0)\log^4\varphi_0+\varphi_0^2(3-6\log\varphi_0+2\log^2\varphi_0+4\log^3\varphi_0)\right]\ ,
\ee
and the bounce profile can be obtained explicitly as
\be
\varphi_B(r)=e^{-\sqrt{r^2/3+\log^2\varphi_0}}\ .
\ee
Again, this example does not admit a thin-wall limit, since the tunneling potential does not feature a 
broken minimum.

\subsection{Example E: Finite mass with thin-wall limit}
Although the last two examples do not admit a thin-wall limit,
this is easily remedied by constructing a second local minimum in $V_t$. For this purpose, take
\be
V_t'(\varphi)=\frac{\varphi (\varphi-1) }{1-\log\varphi}\ ,
\ee
which is finite in $[0,1]$ and has local minima at $\varphi=0$ and $\varphi=1$.
This function integrates to
\be
V_t = e^2\ {\rm Ei}(-2 + 2 \log\varphi) - e^3\ {\rm Ei}(-3 + 3 \log\varphi)\, ,
\ee
and we finally get
\be
V(\varphi)=e^2\ {\rm Ei}(-2 + 2 \log\varphi) - e^3\ {\rm Ei}(-3 + 3 \log\varphi)+\frac{(1-\varphi)^2\varphi^2}{18(1-\log\varphi)^2}r^2(\varphi)\ ,
\ee
with the bounce profile given implicitly by
\be
r^2(\varphi)=3\left[ 2\log \frac{(1 - \varphi) \varphi_0}{(1 - \varphi_0)\varphi}+\log^2\varphi-\log^2\varphi_0  +2 {\rm Li}_2(1 - \varphi) -2 {\rm Li}_2(1 - \varphi_0)\right]\ .
\ee
The thin-wall limit is obtained 
by choosing $\varphi_0$ close to the second minimum, $\varphi_0 \to 1$.


\section{From Single-Field Solution to Two-Field Solutions\label{sec:1to2}} 

\subsection{Paths via curvature and transverse coordinate\label{ss:M1}}

Consider a multi-field potential with a false vacuum at $\phi_i=0$,
with $i=1,...,N$. A tunneling $V_t$ solution satisfies two equations \cite{EK}. One is the scalar single-field equation (\ref{EoM}) where $\varphi$ is now the arc-length along the decay trajectory, with 
\be
d\varphi^2=\sum_{i=1}^N d\phi_i^2 ,
\ee 
and $\varphi=0$ at the false vacuum. The second (transverse and vectorial) equation reads
\be
2(V-V_t){\bma\phi}''={\bma\nabla}_TV\ ,
\label{eq:transverse}
\ee
where ${\bma \phi}=\{\phi_1,\phi_2,...,\phi_N\}$, primes denote as before derivatives with respect to $\varphi$, and
\be
{\bma\nabla}_TV \equiv {\bma \nabla}V-V'{\bma \phi}'\ .
\ee
Equation (\ref{eq:transverse}) shows that the curvature of the decay path is determined by the gradient of the potential along the direction orthogonal to the path.
In the two-field case, this latter equation reduces to a scalar equation that can be written as
\be
2(V-V_t)\theta'=-\partial V/\partial \phi_T\ ,
\label{EoMT2}
\ee
where $\theta(\varphi)$ is the angle between ${\bma\phi}'$ and $\phi_1$ at point $\varphi$ and $\phi_T$ is a (properly normalized) field variable transverse to the ${\bma\phi}$ trajectory. In other words, we have
\be
{\bma\phi}'=(\cos\theta,\sin\theta)\ ,\quad
{\bma\phi}''=(-\sin\theta,\cos\theta)\theta'\ ,
\ee 
and $\theta'$ is nothing but the curvature of ${\bma\phi}$.

Now we assume that  we know $V$ and $V_t$ along the decay trajectory, just by solving a single-field problem, as done in the previous section. Such single-field decay can be associated to an infinite number of two-field potentials. First, note that the path function $\theta(\varphi)$ is totally undetermined by the single-field solution and
we can choose it at will. Second, as  (\ref{EoMT2}) holds only at the  decay path, we have much freedom in integrating it to extend $V$ along directions transverse to the decay path.  A convenient family of solutions is
\be
V_2(\varphi,\phi_T)=V(\varphi) -2f(\phi_T)[V(\varphi)-V_t(\varphi)]\theta'(\varphi)+g(\phi_T,\varphi)\ ,
\label{V2}
\ee
where the functions $f(\phi_T)$ and $g(\phi_T,\varphi)$ satisfy $f(\phi_T)=\phi_T+{\cal O}(\phi_T^2)$ and $g(\phi_T,\varphi)=\phi_T^2g_0(\varphi)+{\cal O}(\phi_T^3)$. We have used the notation $V_2$ to stress that this is a two-field potential.

The previous procedure shows the key elements required to get a two-field example but often it is difficult to find a system of $(\varphi,\phi_T)$ coordinates that covers the  two-field plane (or a large enough part of it) in a consistent way. Below we present two methods that can be used to avoid that problem. 

\subsection{Paths via injective mappings\label{ss:M2}}

In the first method, we start from a path that can be parametrized
unambiguously by one field. 
That is, the path contains every value of e.g.~$\phi_1$ at most once.
Consider a function $\Phi_2(\phi_1)$ that defines the vacuum decay
path via 
\be
(\phi_1,\phi_2) \in (\phi_1,\Phi_2(\phi_1)) \, .
\ee
Given a tunneling potential $V_t(\varphi)$, one can construct the 
potential along the path $V(\varphi)$ using~(\ref{eq:sol_by_int}). A useful ansatz for the  two-field potential in the full parameter space is then given by 
\be
V_2(\phi_1,\phi_2) = V(\varphi(\phi_1)) + W(\phi_1) \left[\phi_2 - \Phi_2(\phi_1)\right] + {\cal O}( [\phi_2 - \Phi_2(\phi_1)]^2) \, .
\label{V2ansatz}
\ee
This involves expressing the arc length along the path, $\varphi$, as a function of $\phi_1$. By definition of the arc length one has
\be
\left(\frac{d\varphi}{d\phi_1}\right)^2=1+\left(\frac{d\Phi_2}{d\phi_1}\right)^2\ .
\ee
The function
$W(\phi_1)$ is matched using the transverse EoM (\ref{eq:transverse}). One gets
\be
W(\phi_1)=V'(\varphi(\phi_1))\frac{\Phi_2'(\phi_1)}{\varphi'(\phi_1)}+2[V(\varphi(\phi_1))-V_t(\varphi(\phi_1))]\left[\frac{\Phi_2''(\phi_1)}{\varphi'(\phi_1)^2}-
\frac{\Phi_2'(\phi_1)\varphi''(\phi_1)}{\varphi'(\phi_1)^3}
\right]
\ ,
\label{W}
\ee
which in this two-field case can be further simplified using
\be
\frac{\Phi_2''(\phi_1)}{\varphi'(\phi_1)^2}-
\frac{\Phi_2'(\phi_1)\varphi''(\phi_1)}{\varphi'(\phi_1)^3}=\frac{\Phi_2''(\phi_1)}{\varphi'(\phi_1)^4}\ .
\ee
The quadratic and higher orders 
in (\ref{V2ansatz}) can be used to make the potential bounded from below. To explain why this is important, consider what happens in a thin-wall case. In such case, both endpoints of the decay trajectory are the two vacua of the potential and one can calculate explicitly the matrix of second derivatives of $V_2$ at these endpoints, either at $\varphi=0$ (false vacuum) or at $\varphi=\varphi_0$ (true vacuum). Without ${\cal O}( [\phi_2 - \Phi_2(\phi_1)]^2)$ terms in $V_2$, one would get the two mass eigenvalues $m_L^2=V''>0$ (corresponding to the field direction tangent to the decay path) and $m_T^2=-(d\Phi_2/d\phi_1)^2V''<0$ (corresponding to the field direction transverse to the decay path), leading to a tachyonic instability.

\subsection{Paths via conformal mappings\label{ss:M3}}

The second method uses the power of conformal mappings to obtain orthogonal curvilinear coordinates that cover the plane. As an example,  from the complex function $F(z)=z^2/2$, writing $z=x+iy$ and $F=u+iv$ we find the relations $u=(x^2-y^2)/2$ and $v=xy$. Lines of constant $u$ and constant $v$ provide such a system of orthogonal coordinates covering the $(x,y)$ plane. This can be used for our purpose by selecting {\it e.g.}~one of the constant $u$ curves for the decay trajectory, while constant $v$ lines orthogonal to the path give a natural definition for the transverse coordinate. This orthogonal coordinate is then quite useful to add in a simple way the quadratic corrections on top of the decay path as discussed above. In the next section we  provide a concrete example realizing this construction.

\section{Two-Field Examples\label{sec:2FieldEx}} 

For the first example of a two-field solution we take example A of one-field solutions described in Subsect.~\ref{sec:ExPolynomial} and, following the  method of Subsect.~\ref{ss:M1}, we  choose $\theta'(\varphi)=1$ so that the decay trajectory is the arc of a  circle in the $(\phi_1,\phi_2)$ plane. We define the transverse field $\phi_T$ by
\be
\phi_1=(1+\phi_T)\sin\varphi\ , \quad \phi_2=1-(1+\phi_T)\cos\varphi\ ,
\ee
so that the decay trajectory corresponds to $\phi_T=0$.
We then have
\be
\varphi(\phi_1,\phi_2)=\arctan\left(\frac{1-\phi_2}{\phi_T+1},\frac{\phi_1}{\phi_T+1}\right)\ ,\quad
\phi_T(\phi_1,\phi_2) =\sqrt{\phi_1^2+(1-\phi_2)^2}-1\ .
\ee
Selecting next the functions $f$ and $g$ in (\ref{V2}) as
\be
f(\phi_T)=\phi_T-20\phi_T^2\ ,\quad g(\phi_T,\varphi)=\phi_T^2\ ,
\label{paramchoice}
\ee
we have all the ingredients needed for a two-field potential.

\begin{figure}[t!]
\begin{center}
\includegraphics[width=0.4\textwidth]{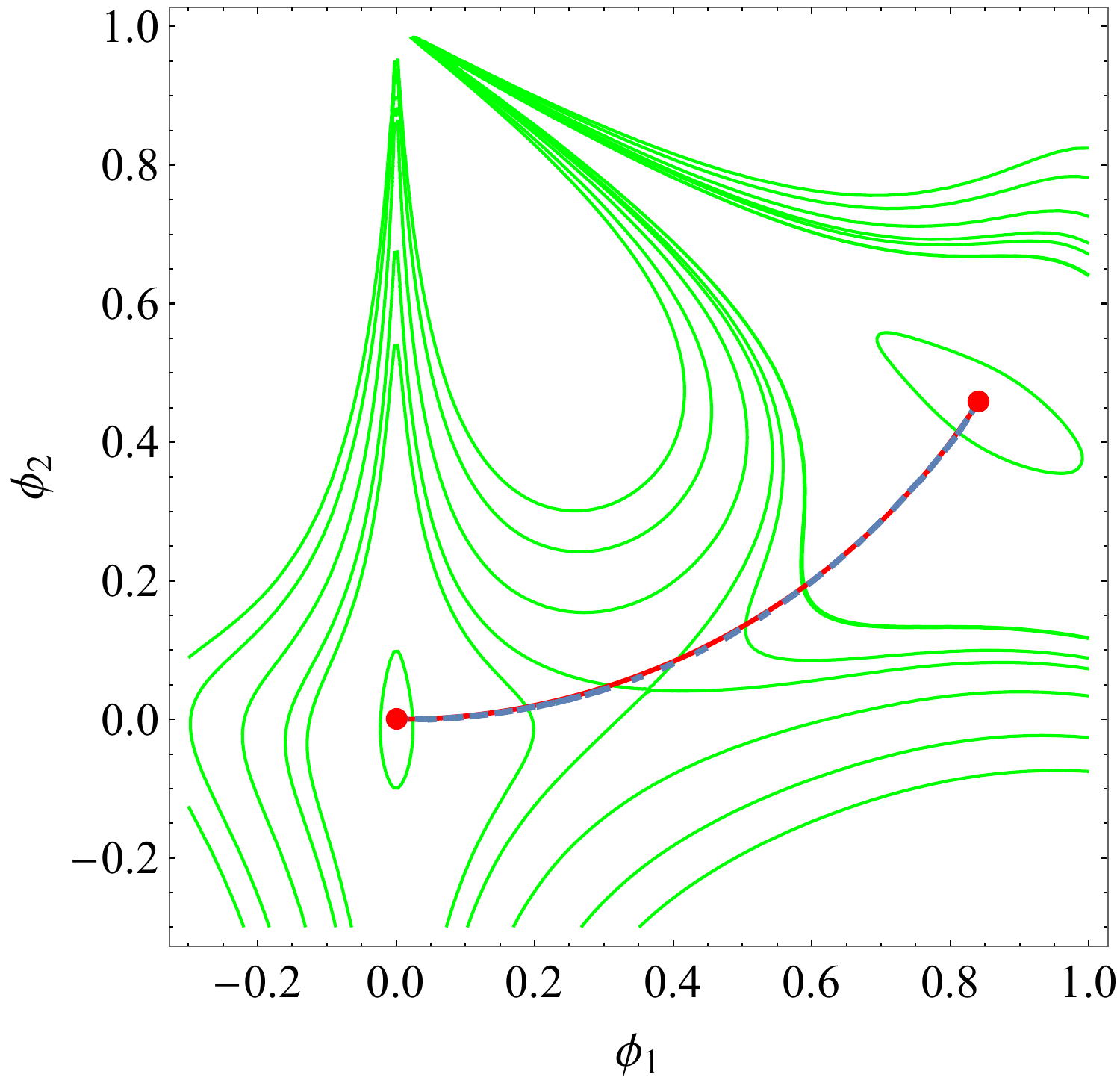}\hspace{0.5cm}
\includegraphics[width=0.45\textwidth]{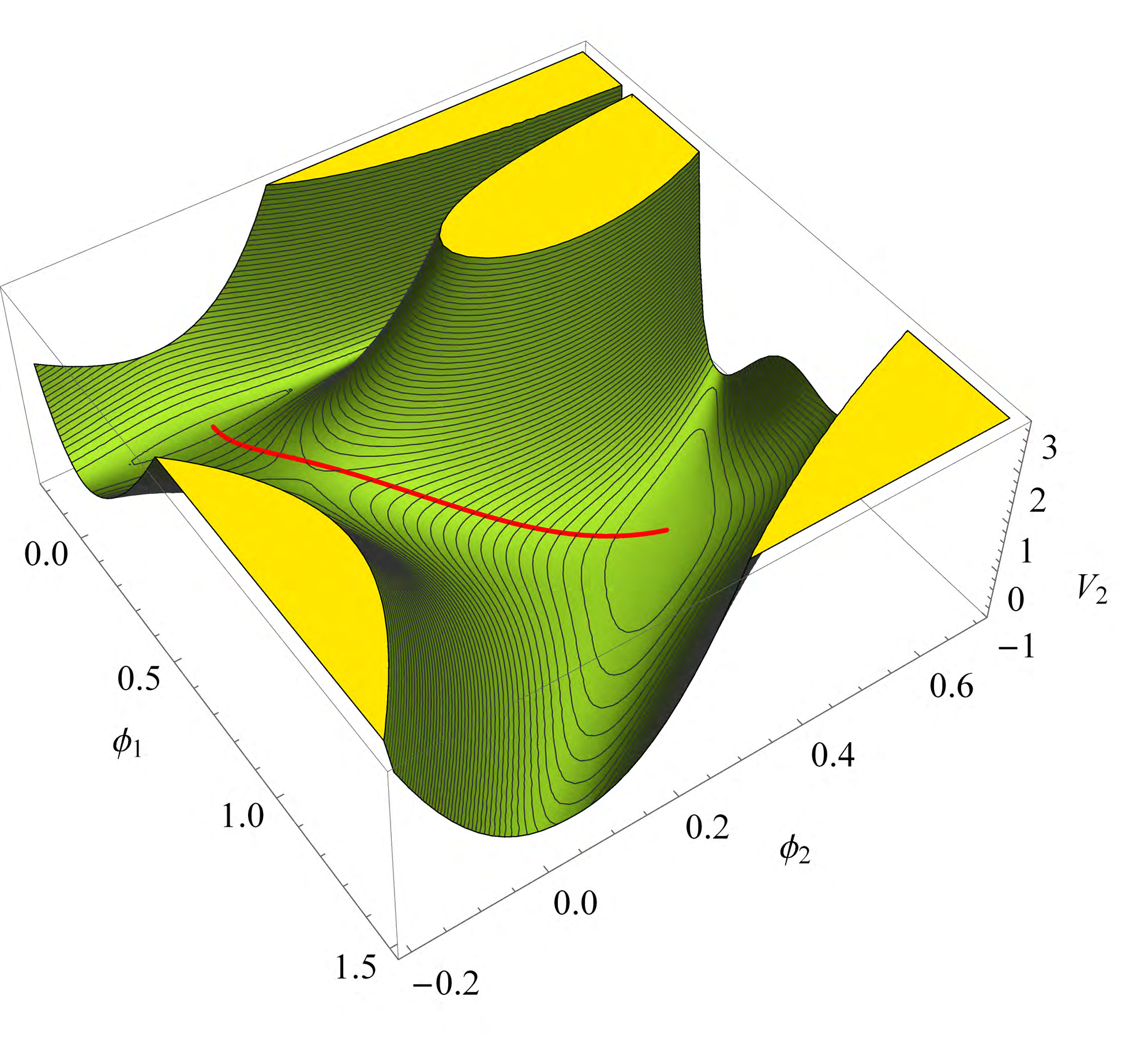}
\end{center}
\caption{For the potential of eq.~(\ref{V2}) with $\theta'(\varphi)=1$, the functions $f$ and $g$ chosen as in (\ref{paramchoice}) and $\varphi_0=0.999$, the left plot shows contour lines of $V_2$  (green) and the decay path between minima both analytically (red) and as found by the FindBounce \cite{FB} package (blue-dashed). The right plot shows a 3D version of the same.
\label{fig:2field}
}
\end{figure}

In figure~\ref{fig:2field} we show the two-field potential as well as the decay trajectory, for the choice $\varphi_0=0.999$ (corresponding to a bounce radius $R=2.63$). In the contour plot we compare the analytical trajectory (red curve) with the trajectory found using the public code FindBounce \cite{FB} (blue dashed curve). The numerical action is computed to be $S=80.70$ while the analytical result is $S=80.59$.
As already mentioned, we can choose other $\theta,f,g$ functions to generate other examples.  The particular choice we made helps to clarify the difficulties that can be encountered in trying to find a global definition of $\phi_T$. In this example, the point $(\phi_1=0,\phi_2=1)$ 
corresponds to $\phi_T=-1$ and any value of $\varphi$, and this implies that the potential $V_2(\phi_1,\phi_2)$ is undefined at that point. One could ignore the region of field space near that point, or regularize the potential by writing {\it e.g.}
\be
V_2(\varphi,\phi_T)=(1-\phi_T^3)\left\{V(\varphi) -2\phi_T[V(\varphi)-V_t(\varphi)]\right\}+g_0\phi_T^2\ .
\ee
Nevertheless, it is simpler to use the method introduced in Subsect. \ref{ss:M2}, which we do next.

As a simple example, consider the decay path $\Phi_2(\phi_1) = \alpha \cosh(\phi_1/\alpha)-\alpha$. This leads to 
$d\varphi = \cosh(\phi_1/\alpha) \, d\phi_1$ and hence $\varphi = \alpha \sinh(\phi_1/\alpha)$ [we set the false minimum at $(0,0)$ and $\varphi=0$ at it]. Parametrized by $\varphi$, the decay path is 
\be
(\phi_1,\phi_2) \in (\alpha \,  \textrm{arcsinh}(\varphi/\alpha) ,  \sqrt{\alpha^2 + \varphi^2}-\alpha )
\ee
Hence
\be
{\bma\phi}^\prime = \frac{1}{\sqrt{\alpha^2 + \phi^2}}\left( 
\alpha,\phi\right) \ , \quad\quad
{\bma\phi}^{\prime\prime} =\frac{\alpha}{(\alpha^2 + \phi^2)^{3/2}} \left( 
-\phi,\alpha \right)
\ee
For this path we get, from (\ref{W}),
\be
W(\phi_1) =V' \tanh(\phi_1/\alpha) +\frac{2(V-V_t)}{\alpha\cosh(\phi_1/\alpha)^3}\ ,
\ee
with $V$ and $V_t$ considered as functions of $\varphi$ and the function $\varphi(\phi_1)$ as given above.

\begin{figure}[t!]
\begin{center}
\includegraphics[width=0.4\textwidth]{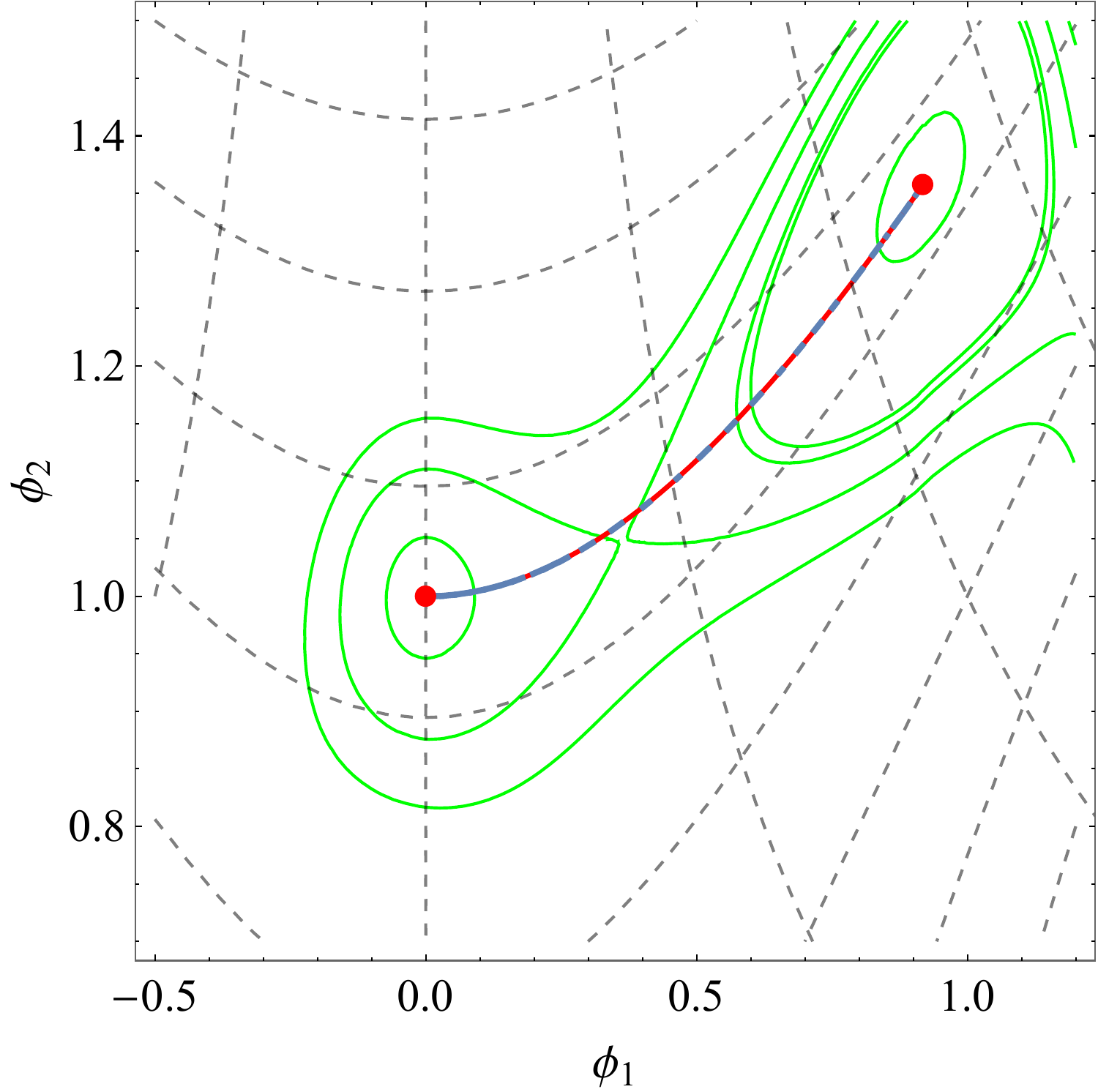}
\hspace{1cm}
\includegraphics[width=0.45\textwidth]{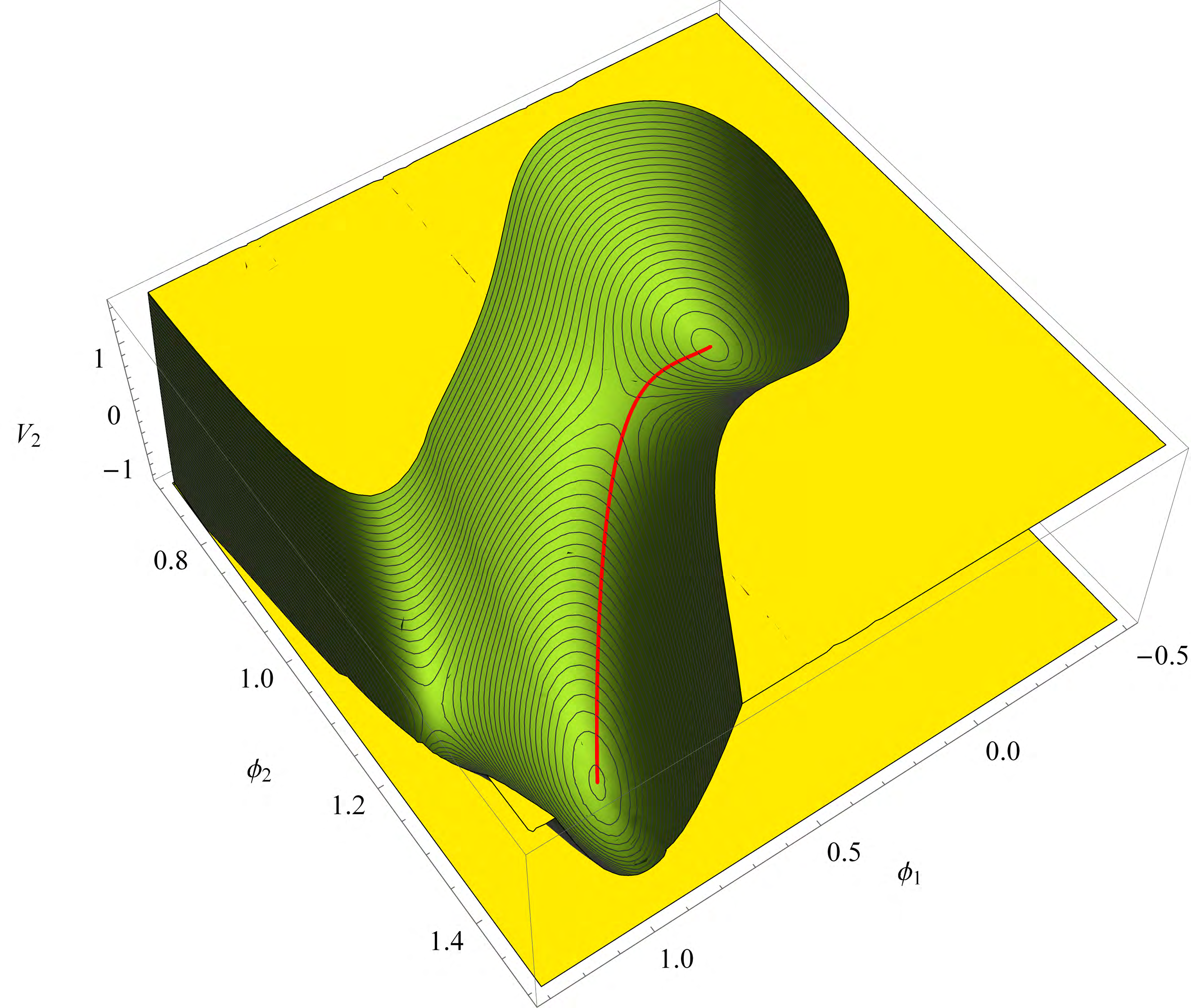}
\end{center}
\caption{For the potential of (\ref{V2Conf}) with the parameters chosen as explained in the text, the left plot shows contour lines of $V_2$  (green) and the decay path between minima both analytically (red) and as found by the FindBounce \cite{FB} package (blue-dashed). The right plot shows a 3D version of the same.
\label{fig:Conf}
}
\end{figure}

Finally, to illustrate the use of conformal mappings as a tool to get orthogonal curvilinear coordinates, we take, as in 
Subsect.~\ref{ss:M3}, $F(z)=z^2/2$, giving $u+iv=(x^2-y^2)/2+ixy$.  Identifying $x=\phi_1$ and $y=\phi_2$, the lines of constant $u$ and $v$ give a system of orthogonal curvilinear coordinates in the $(\phi_1,\phi_2)$ plane.  For our example we take the decay path to lie in the line $u=-1/2$, with 
\be
\Phi_2(\phi_1)=\sqrt{1+\phi_1^2}\ .
\ee
Figure~\ref{fig:Conf} shows the path (blue line) and the system of coordinates (dashed black lines). From $\Phi_2(\phi_1)$ above one gets the arc length 
\be
\varphi(\phi_1)=-i\, {\rm E}[i {\rm Arcsinh}(\phi_1),2]\ ,
\ee
where $E[\phi,m]$ is the incomplete elliptic function of the second kind. We then write the two-field potential as
\be
V_2(\phi_1,\phi_2)=V(\varphi(\phi_1))+W(\phi_1)[\phi_2-\Phi_2(\phi_1)]+\frac12 M_T^2(\phi_1,\phi_2)(\phi_2^2-\phi_1^2-1)^2\ ,
\label{V2Conf}\ee
where the second order term is proportional to the square of the transverse distance from the path, with $M_T^2>0$ large enough to avoid instabilities of the potential.

In Figure~\ref{fig:Conf} we give such an example, for the same single-field potential used above and taking $M_T^2=9(2-\phi_1)$,
which is positive in the region of interest. The action obtained numerically with the FindBound package is $S\simeq 80.66$ (while the exact one is $S=80.59$).

\section{From Single-Field Solution to Three-Field Solutions \label{sec:1to3}} 

Just as in the case with two fields, one can construct solutions with three fields 
using a path that is injective in one variable. (In fact, the following construction can be trivially extended to any number of fields.) 
Writing the decay path as 
\be
(\phi_1,\phi_2,\phi_3)\in (\phi_1,\Phi_2(\phi_1),\Phi_3(\phi_1))\ ,
\ee 
with arc length field $\varphi$ satisfying
\be
\left(\frac{d\varphi}{d\phi_1}\right)^2=1+\left(\frac{d\Phi_2}{d\phi_1}\right)^2+\left(\frac{d\Phi_3}{d\phi_1}\right)^2\ ,
\ee
we take
\be
V_3(\phi_1,\phi_2,\phi_3)=V+W_2(\phi)(\phi_2-\Phi_2(\phi_1))+W_3(\phi)(\phi_3-\Phi_3(\phi_1))+...
\ee
where the dots stands for higher order terms that stabilize the trajectory.
The functions $W_{2,3}(\phi_1)$ are obtained from the transverse EoM (\ref{eq:transverse}). One gets
\bea
W_2(\phi_1)&=&V'(\varphi(\phi_1))\frac{\Phi_2'(\phi_1)}{\varphi'(\phi_1)}+2[V(\varphi(\phi_1))-V_t(\varphi(\phi_1))]
\left[
\frac{\Phi_2''(\phi_1)}{\varphi'(\phi_1)^2}-\frac{\Phi_2'(\phi_1)\varphi''(\phi_1)}{\varphi'(\phi_1)^3}\right] ,\nonumber\\
W_3(\phi_1)&=&V'(\varphi(\phi_1))\frac{\Phi_3'(\phi_1)}{\varphi'(\phi_1)}+2[V(\varphi(\phi_1))-V_t(\varphi(\phi_1))]\left[\frac{\Phi_3''(\phi_1)}{\varphi'(\phi_1)^2}-\frac{\Phi_3'(\phi_1)\varphi''(\phi_1)}{\varphi'(\phi_1)^3}\right] ,
\label{W2}
\eea
This construction is quite straightforward and can be further simplified 
by choosing an appropriate path. This is demonstrated in the example of the next section.

\section{Three-Field Example\label{sec:3FieldEx}} 

\begin{figure}[t!]
\begin{center}
\includegraphics[width=0.6\textwidth]{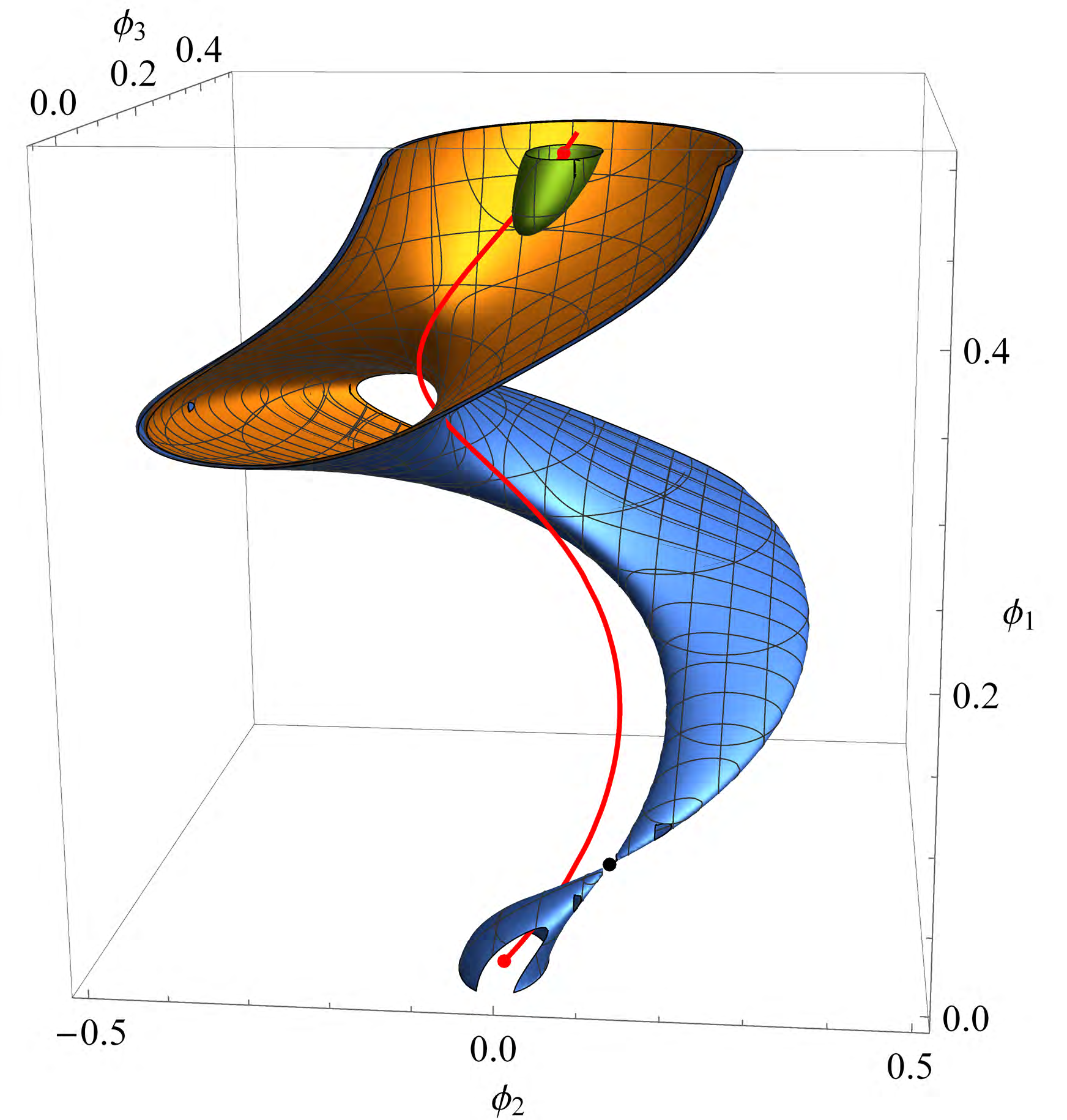}
\end{center}
\caption{
\label{fig:3field}
Equipotential surfaces for the three-field potential of (\ref{V3}). The red line shows the decay trajectory between the red points of false vacuum, at $(0,0,0)$, and tunneling endpoint. The black dot is a potential saddle point. 
}
\end{figure}

To construct a 3D example we first select a decay trajectory. The helix is a particularly interesting example
\be
(\phi_1,\phi_2,\phi_3)\in \left(\alpha\varphi,\rho\sin(\sqrt{1-\alpha^2}\varphi/\rho),\rho-\rho\cos(\sqrt{1-\alpha^2}\varphi/\rho)\right)\ ,
\ee
where $\varphi$ is the arc length along the path. The winding of the helix typically precludes 
a fully numerical search for the bounce, while the construction is simple enough to track 
the solution analytically. 
We can use $\phi_1$ to parametrize this path and write
\be
\Phi_2(\phi_1)=\rho \sin\left(\frac{\sqrt{1-\alpha^2}}{\rho\alpha}\phi_1\right)\ ,\quad
\Phi_3(\phi_1)=\rho-\rho \cos\left(\frac{\sqrt{1-\alpha^2}}{\rho\alpha}\phi_1\right)\ ,
\ee
as well as $\varphi(\phi_1)=\phi_1/\alpha$. We then define the three-field potential as
\bea
V_3(\phi_1,\phi_2,\phi_3)&=&V(\varphi(\phi_1))+W_2(\phi_1)(\phi_2-\Phi_2(\phi_1))+W_3(\phi_1)(\phi_3-\Phi_3(\phi_1))\nonumber\\&&+25(\phi_2-\Phi_2(\phi_1))^2+25(\phi_3-\Phi_3(\phi_1))^2\ ,
\label{V3}
\eea
where $V$ is a single-field potential like those discussed in Section~\ref{sec:field} and $W_{2,3}(\phi_1)$ have been defined in the previous section.

For our numerical example we take $\alpha=1/2$, $\rho=1/8$ and use the potential $V$ of example~A in Subsection~\ref{sec:ExPolynomial}, with $\varphi_0=0.999$ (a thin-wall case). Figure
\ref{fig:3field} shows equipotential surfaces of $V_3(\phi_1,\phi_2,\phi_3)$ and the decay trajectory (red line).

\section{Including Gravity\label{sec:Gravity}} 
Gravity can change qualitatively vacuum decay \cite{CdL} and such effects can be described in a simple way by the tunneling potential method, see \cite{Eg}. The tunneling action is now given by
\be
S[V_t]=\frac{6\pi^2}{\kappa^2}\int_{\varphi_+}^{\varphi_0}\frac{(D+V_t')^2}{DV_t^2}\ d \varphi\ ,
\ee
with $\kappa=1/m_P^2$, where $m_P$ is the reduced Planck mass, and
\be
D\equiv \sqrt{V_t'{}^2+6\kappa(V-V_t)V_t}\ .
\ee
The Euler-Lagrange equation for $V_t$ is
\be
(4V_t'-3V')V_t'+6(V-V_t)\left[V_t''+\kappa(3V-2V_t)\right]=0\ .
\label{EoMVtg}
\ee
The $V_t$ method is also useful to generate exact examples of vacuum decay including gravitational corrections. We review here how this works, see \cite{Eg,EFH} for more details.

Starting from a simple enough $V_t(\varphi)$ we construct $V(\varphi)$ as
\be
V(\varphi)= V_t+\frac{V_t'{}^2}{6\kappa(1/F-V_t)}\ ,
\ee
where the function $F$ is given by
\be
F(\varphi)=\frac{2\kappa}{E(\varphi)}\int_{\varphi_0}^{\varphi}
\frac{E(\tilde \varphi)}{V_t'(\tilde\varphi)}d\tilde\varphi\ , \quad
E(\varphi)={\rm exp}\left[2\kappa\int_{\varphi_0}^{\varphi}
\frac{V_t(\tilde \varphi)}{V_t'(\tilde\varphi)}d\tilde\varphi\right] \ .
\ee
Alternatively \cite{EFH}, one can try to get $F$ directly by solving
\be
F'V_t'=2\kappa(1-F V_t)\ .
\ee 
In those cases that admit an exact solution of the integrals for $E$ and $F$ above, one can always obtain the related metric function $\rho(\varphi)$.\footnote{The field profile of the Euclidean bounce can be obtained in some cases by integrating
the relation $d\varphi/d\xi=-\sqrt{2(V-V_t)}$, where $\xi$ is the radial coordinate in the Euclidean metric $ds^2=d\xi^2+\rho^2(\xi)d\Omega_3^2$, where $d\Omega_3^2$ is the line element of the unit 3-sphere. The metric function $\rho(\xi)$ can also be recovered from $V_t$ using the relation $\rho=\sqrt{2(V-V_t)}/D$.} In terms of this function the potential reads
\be
V(\varphi)=V_t(\varphi)+\frac{1}{18}D^2(\varphi)\rho^2(\varphi)\ ,
\ee
which reduces to (\ref{Vr2}) for $\kappa=0$.

Next, we provide an example of potential for which vacuum decay can be treated analytically using the method just discussed. (Previous exact examples, found using different approaches, can be found in \cite{DH,exactgrav}.) Let us take \cite{Eg}
\be
V_t = V_+ -\sin^2\varphi\ ,
\label{Vtan}
\ee
with $V_+\leq 0$ (Minkowski or AdS vacua). Setting $\kappa=1$, one gets
\be
V = V_t -\frac{s_{2\varphi}^2}{6 V_t}\left\{
1+\frac{c_\varphi^{\alpha}s_\varphi^{-(1+\alpha)}}{2[A(\varphi)+C]V_t}\right\}\ ,
\label{Vanalytic}
\ee
where $s_\varphi\equiv \sin\varphi$, $c_\varphi\equiv \cos\varphi$, $\alpha\equiv V_+-1$. The function $A(\varphi)$ is given in terms of the Appell hypergeometric function of two variables, $F_1(a;b_1,b_2;c;x,y)$, as
\be
A(\varphi)\equiv
\frac{c_\varphi^{2+\alpha}}{\alpha^2(2+\alpha)}
F_1\left(a;b_1,b_2;c;c_\varphi^2;-c_\varphi^2/\alpha\right)\ ,
\ee
with $a=1+\alpha/2$, $b_1=(1+\alpha)/2$, $b_2=2$, and $c=2+\alpha/2$.
Finally, $C$ is an integration constant that can be expressed in terms
of $\phi_0$ solving $V(\phi_0)=V_t(\phi_0)$. 

The Minkowski limit ($V_+=0$) of the previous result can be expressed in terms of elementary functions and reads
\be
V=-s_\varphi^2+\frac23 c_\varphi\left\{c_\varphi-\frac{1}{s_\varphi^2\left[{\rm arctanh} (c_\varphi)-{\rm arctanh} (c_{\varphi_0})+1/c_{\varphi_0}\right]+c_\varphi}\right\}\ .
\ee
Additional examples for false dS vacua, obtained using the same technique, can be found in \cite{EFH}.

\section{Conclusions and Outlook\label{sec:out}} 

We have constructed various exact solutions to the tunneling problem with several scalar fields. Our method to generate such examples relies on exact solutions for a single scalar field uplifted to more field dimensions along an almost arbitrary path in scalar multi-field space. Our examples and general technique can be used for a number of applications as has been the case in the past (see a partial list in the introduction) opening the scope to multifield problems.

We leave possible applications of our techniques  for future work, but comment
briefly on the use as benchmarks for the numerical codes that are frequently used for phenomenological analyses of vacuum decay \cite{VEVACIOUS,CT,AB,BP,SB,FB}.
Numerical codes often struggle with multi-field setups, in particular
if a good first guess for the path cannot be provided. We tried 
several public codes on the three-field helix example of Section \ref{sec:3FieldEx} and all of them failed except CosmoTransitions \cite{CT}, which was able to converge to the correct path but only after providing a good starting path for the algorithm. 
The final action was correct up to a few per-mille with about a hundred support points for the path.

We also provided some exact non-trivial examples for tunneling 
including gravity. Public codes to solve this problem numerically are 
not yet available and the exact examples provided here should be helpful for their development. 

\section*{Acknowledgments\label{sec:ack}} 

The work of J.R.E.~has been funded by the following grants: IFT Centro de Excelencia Severo Ochoa CEX2020-001007-S and by PID2022-142545NB-C22 funded by MCIN/AEI/10.13039/ 501100011033 and by ``ERDF A way of making Europe''.
T.K.~acknowledges support by the Deutsche Forschungsgemeinschaft (DFG, German Research Foundation) under Germany’s Excellence Strategy – EXC 2121 ``Quantum Universe'' – 390833306.

\end{document}